\documentstyle{article}
\input math_macros
\font\tenbf=cmbx10

\font\tenrm=cmr10
\font\tenit=cmti10
\font\elevenbf=cmbx10 scaled\magstep 1
\font\elevenrm=cmr10 scaled\magstep 1
\font\elevenit=cmti10 scaled\magstep 1

\font\ninerm=cmr9

\def \cn{Collaboration}
\def \cp89{{\it CP Violation,} edited by C. Jarlskog (World Scientific,
Singapore, 1989)}

\def \hb87{{\it Proceeding of the 1987 International Symposium on Lepton and
Photon Interactions at High Energies,} Hamburg, 1987, ed. by W. Bartel
and R. R\"uckl (Nucl. Phys. B, Proc. Suppl., vol. 3) (North-Holland,
Amsterdam, 1988)}

\def \ichep72{{\it Proceedings of the XVI International Conference on High
Energy Physics}, Chicago and Batavia, Illinois, Sept. 6 -- 13, 1972,
edited by J. D. Jackson, A. Roberts, and R. Donaldson (Fermilab, Batavia,
IL, 1972)}

\def \ite{{\elevenit et al.}}
\def \lkl87{{\it Selected Topics in Electroweak Interactions} (Proceedings of
the Second Lake Louise Institute on New Frontiers in Particle Physics, 15 --
21 February, 1987), edited by J. M. Cameron \ite~(World Scientific, Singapore,
1987)}
\def \ky85{{\it Proceedings of the International Symposium on Lepton and
Photon Interactions at High Energy,} Kyoto, Aug.~19-24, 1985, edited by M.
Konuma and K. Takahashi (Kyoto Univ., Kyoto, 1985)}

\def \si90{25th International Conference on High Energy Physics, Singapore,
Aug. 2-8, 1990}
\def \slc87{{\it Proceedings of the Salt Lake City Meeting} (Division of
Particles and Fields, American Physical Society, Salt Lake City, Utah, 1987),
ed. by C. DeTar and J. S. Ball (World Scientific, Singapore, 1987)}
\def \slac89{{\it Proceedings of the XIVth International Symposium on
Lepton and Photon Interactions,} Stanford, California, 1989, edited by M.
Riordan (World Scientific, Singapore, 1990)}
\def \smass82{{\it Proceedings of the 1982 DPF Summer Study on Elementary
Particle Physics and Future Facilities}, Snowmass, Colorado, edited by R.
Donaldson, R. Gustafson, and F. Paige (World Scientific, Singapore, 1982)}
\def \smass90{{\it Research Directions for the Decade} (Proceedings of the
1990 Summer Study on High Energy Physics, June 25--July 13, Snowmass,
Colorado),
edited by E. L. Berger (World Scientific, Singapore, 1992)}
\def \tasi90{{\it Testing the Standard Model} (Proceedings of the 1990
Theoretical Advanced Study Institute in Elementary Particle Physics, Boulder,
Colorado, 3--27 June, 1990), edited by M. Cveti\v{c} and P. Langacker
(World Scientific, Singapore, 1991)}

\renewenvironment{thebibliography}[1]
 { \elevenrm
   \begin{list}{\arabic{enumi}.}
    {\usecounter{enumi} \setlength{\parsep}{0pt}
     \setlength{\itemsep}{3pt} \settowidth{\labelwidth}{#1.}
     \sloppy
    }}{\end{list}}

\begin{document}
\begin{center}{{\tenbf HEAVY MESON MASSES AND DECAY CONSTANTS\footnote{
\ninerm Presented at DPF 92 Meeting, Fermilab, November, 1992.}\\}
\vspace{-1in}
\rightline{EFI 92-61}
\rightline{November 1992}
\bigskip
\vglue 2.0cm
{\tenrm JONATHAN L. ROSNER\\}
\baselineskip=13pt
{\tenit Enrico Fermi Institute and Department of Physics,
University of Chicago\\}
\baselineskip=12pt
{\tenit 5640 S. Ellis Ave., Chicago, IL 60637, USA\\}
\vglue 0.8cm
{\tenrm ABSTRACT}}
\end{center}
\vglue 0.3cm
{\rightskip=3pc
 \leftskip=3pc
 \tenrm\baselineskip=12pt
 \noindent
Masses and decay constants of mesons containing a single $c$ or $b$ quark are
described within the framework of heavy-quark symmetry.  The $B_s^* - B_s$ and
$\bar B^{*0} - \bar B^0$ mass splittings are found equal to within a fraction
of an MeV. Decay constants of $D$ and $B$ mesons are estimated using isospin
mass splittings in the $D$, $D^*$, $B$, and $B^*$ states to isolate the
electromagnetic hyperfine interaction between quarks.  A relation following
from the use of splittings in kaons is also considered.
\vglue 0.6cm}
{\elevenbf\noindent 1. Introduction}
\vglue 0.4cm
\baselineskip=14pt
\elevenrm

Mesons containing one heavy quark $(c,b)$ are of fundamental importance for the
understanding of the strong interactions, since they consist of a single light
quark bound to a nearly static source of color.  We describe here some recent
work on the masses \cite{RW} and decay constants \cite{FDB} of such mesons.

Results on heavy meson masses come from an expansion to first order in
$\alpha$, first order in light-quark masses $(m_u,~m_d,~m_s)$, and first order
in $1/m_Q$, where $Q$ is a heavy quark.  We predict one new relation:  The
photons in $B_s^* \to B_s \gamma$ and $B^{*0} \to B^0 \gamma$ should have equal
energies.

Decay constants of heavy mesons are crucial for interpreting data on
particle-antiparticle mixing in the neutral $B$ meson system, and for
anticipating and interpreting new signatures for CP violation.  We describe a
method for determination of these constants which relies on the isospin
splittings of the $D$, $D^*$, $B$, and $B^*$ mesons. We also consider a
relation following from the use of splittings in kaons. The isospin splittings
allow one to extract the contributions of the spin-dependent electromagnetic
interaction between light and heavy quarks. Additional assumptions about quark
masses are required in order to interpret these contributions in terms of decay
constants.
\vglue 0.5cm
{\elevenbf\noindent 2. Masses of $D$ and $B$ mesons}
\vglue 0.2cm
The most general mass operator containing contributions of first order in (a)
light quark masses $m_q$, (b) electromagnetic interactions, and (c) $1/m_Q$,
including terms of order $m_q/m_Q$ and $\alpha/m_Q$, contains 11 terms, not
counting ones which can be absorbed into heavy quark masses \cite{RW}. One
result of this expansion is the familiar relation between the strong hyperfine
splitting between the $^3S_1$ and $^1S_0$ $D$ and $B$ states, which says that
$\Delta M^2$ should be approximately the same for the two systems.  Aside from
this result, we find one new prediction:
\begin{equation}
[B_s^* - B_s] - [\bar B^{*0} - \bar B^0]
= (m_c/m_b)( [D_s^* - D_s] - [D^{*+} - D^+] )~~~,
\end{equation}
where here and below symbols stand for particle masses.  Since $D_s^* - D_s =
141.5 \pm 1.9 {\rm~MeV}$ \cite{PDG} and $D^{*+} - D^+ = 140.64 \pm 0.08 \pm
0.06 {\rm~MeV}$ \cite{B+}, we expect the right-hand side of this relation to be
about $0.3 \pm 0.6 {\rm~MeV}$. A recent study suggests that the smallness of
the result could be due to accidental cancellation, and that there could be
additional contributions of up to a few MeV from effects of higher order in
$1/m_Q$ \cite{Randall}.
\vglue 0.5cm
{\elevenbf \noindent 3. Predictions for decay constants}
\vglue 0.2cm
In the nonrelativistic formula
\begin{equation}
f_M^2 = \frac{12 |\Psi(0)|^2}{M_M^2}~~~
\end{equation}
we seek an estimate of $\Psi(0)$, the nonrelativistic wave function at zero
separation of the light and heavy quark.  This may be obtained in a constituent
quark model from the contribution of electromagnetic hyperfine splitting to
meson masses.  Specifically, in the limit in which the wave functions of a
light quark bound to a $c$ and $b$ quark are the same,
\begin{equation}
\Delta(D) \equiv (D^+ - D^0) - (D^{*+} - D^{*0}) = a + \frac{8 \pi \alpha
Q_c}{3 m_u m_c} |\Psi(0)|^2~~~, \end{equation}
\begin{equation}
\Delta(B) \equiv (\bar B^0 - B^-) - (\bar B^{*0} - B^{*-}) = \frac{m_c}{m_b} a
+ \frac{8 \pi \alpha Q_b}{3 m_u m_b} |\Psi(0)|^2~~~.
\end{equation}
Here $a$ denotes the effects of $m_u \neq m_d$ in the color hyperfine
interaction and of spin-dependent light-quark electromagnetic self-energies.
Now, while we know \cite{B+} that $\Delta(D) = 4.80 \pm 0.11 {\rm~MeV}$, the
corresponding value \cite{DB} for $B$ mesons, $\Delta(B) = 0.12 \pm 0.58
{\rm~MeV}$, is too poorly known to allow us to separate the effects of $a$ and
$|\Psi(0)|^2$. In Fig.~1 we show the dependence of predicted decay contants on
$\Delta(B)$.

In order to proceed further we use a trick motivated by a result of Cohen and
Lipkin \cite{CL} which appeals to the similarity between the kaon and $B$
systems.  We define
\begin{equation}
\Sigma (B) \equiv (\bar B^{*0} + B^{*-}) - (\bar B^0 + B^-)~~~,
\end{equation}
with similar definitions [cf.~(4)] for $\Delta(K)$ and $\Sigma(K)$.  We then
estimate
\begin{equation}
\Delta(B) = \Delta(K) \Sigma(B)/ \Sigma(K) = (-0.06 \pm 0.04)~{\rm MeV}~~~.
\end{equation}
As a result, we can separate out the electromagnetic hyperfine term in (2) and
(3), finding
\begin{equation}
|\Psi(0)|^2 = (13.8 \pm 1.4) \times 10^{-3}~{\rm GeV}^3~~,~~~
f_D^{(0)} = (290 \pm 15)~{\rm MeV}~~,~~~
f_B^{(0)} = (177 \pm  9)~{\rm MeV}~~~.
\end{equation}
\vglue 0.5cm
{\elevenbf \noindent 4. Comparison with experiment}
\vglue 0.4cm
The Mark III Collaboration finds $B(D \to \mu \nu) \times 10^{-4}$ (90\% c.l.),
corresponding to $f_D < 290 {\rm~MeV}$.  The lowest-order result (7) obtained
suggests that $f_D$ may be close to its present upper limit, so a search for $D
\to \mu \nu$ (e.g., through the reaction $e^+ e^- \to \psi(3770) \to D^+ D^-$
at the Beijing Electron Synchrotron) should prove fruitful.

\begin{figure}
\vspace{5in}
\caption{Decay constants $f_D^{(0)}$ (solid curve) and $f_B^{(0)}$ (dashed
curve) predicted by the nonrelativistic formula (2) as functions of difference
$\Delta(B)$ in isospin splittings between $B^*$ and $B$ mesons.  The horizontal
line with the arrow pointing downward denotes the upper limit of Ref.~[7] on
$f_D$.}
\end{figure}

It may be possible, for example at CLEO, to look for the decay $D \to \mu \nu$
by tagging a $D^{\pm}$ using the reaction $D^{*\pm} \to \pi^0 D^{\pm}$, since
the $\pi^0$ is very soft in the $D^*$ or $D$ center-of-mass system, and helps
to label the frame of the decaying $D$.  The signal will show up in a
characteristic band of $m(\pi^0 \mu)$.  One will probably need additional
kinematic information to reduce backgrounds (e.g., from semileptonic decays).

Recent evidence for the reaction $D_s \to \mu \nu$ in emulsion \cite{WA75}
rests on the observation of a muon beyond the kinematic endpoint for
semileptonic decays of $D^+$ and $D_s$.  A search for $D_s \to \mu \nu$ using
the information from the photon in $D_s^* \to D_s \gamma$ is possible in
principle \cite{SSPC}. One needs additional jet or missing energy information.

The WA75 result \cite{WA75}, $f_{D_s} = 232 \pm 69 {\rm~MeV}$ (based on 7
events above background) may be used indirectly to estimate corrections of
order $1/m_Q$ to the lowest-order formula (2).  First we estimate
$|\Psi(0)|_D^2$ using the approximate equality of strong hyperfine splittings
in the $D$ and $D_s$ systems, which implies that
\begin{equation}
|\Psi(0)|_D^2/m_u m_c = |\Psi(0)|_{D_s}^2/m_s m_c~~~.
\end{equation}
With $m_u/m_s = 310 {\rm~MeV}/485 {\rm~MeV}$, we then estimate from the
observed value of $f_{D_s}$ that $f_D = 190 \pm 57 {\rm~MeV}$.  If the
discrepancy with our lowest-order prediction is ascribed to $1/m_Q$
corrections, we may write $f_D = f_D^{(0)}(1 - [\Delta/m_D])$, implying
$\Delta/m_D = 0.35 \pm 0.20$ and hence $\Delta/m_B = 0.13 \pm 0.07$ or $f_B =
f_B^{(0)}(1 - [\Delta/m_B]) = (154 \pm 17) {\rm~MeV}$.  QCD corrections
probably raise this value by about 10\%. The most recent range of lattice gauge
theory values quoted at this conference \cite{Sharpe} puts $f_B$ in the range
between 175 and 200 MeV and $f_D$ just slightly above 200 MeV.  A value of
$f_B$ of about 170 MeV is entirely compatible with recent fits to parameters of
the Cabibbo-Kobayashi-Maskawa matrix and $B - \bar B$ mixing data.
\vglue 0.5cm
{\elevenbf \noindent 5. Acknowledgements \hfil}
\vglue 0.4cm
I thank Jim Amundson, Nahmin Horwitz, Mike Kelly, Sheldon Stone, and Mark Wise
for collaboration on some of the topics mentioned here, and Glenn Boyd and
Harry Lipkin for helpful discussions.  This work was supported in part by the
United States Department of Energy under grant No. DE AC02 90ER40560.
\vglue 0.5cm
{\elevenbf\noindent 6. References \hfil}

\end{document}